\newcommand{\rem}[1]{}
\documentclass{amsart}
\usepackage{amsfonts,amssymb,amsmath,amsthm}
\usepackage{url}
\usepackage{epsfig}

\newtheorem{thrm}{Theorem}[section]

\newtheorem{prop}[thrm]{Proposition}
\newtheorem{cor}[thrm]{Corollary}
\newtheorem{remark}[thrm]{Remark}
\theoremstyle{definition}

\author{Luca~Guido~Molinari and Giuseppe~Lacagnina}
\address{Physics Department\\
Universit\'a degli Studi di Milano and I.N.F.N. sez. Milano\\
Via Celoria 16, 20133 Milano, Italy}
\email{luca.molinari@mi.infn.it}
\keywords{Transfer matrix, block triangular matrix, band matrix, Anderson model,
Lyapunov spectrum.}
\date{Feb 8, 2011}
\begin{document}
\title{Counting the exponents of single transfer matrices}

\begin{abstract}
The eigenvalue equation of a band or a block tridiagonal matrix,
the tight binding model for a crystal, a molecule, or a particle in a
lattice with random potential or hopping amplitudes: these and other 
problems lead to three-term recursive relations for (multicomponent) 
amplitudes.
Amplitudes $n$ steps apart are linearly related by a transfer matrix, 
which is the product of $n$ matrices. 
Its exponents describe the decay lengths of the amplitudes.  
A formula is obtained for the counting function of the exponents, based on
a duality relation and the Argument Principle for the zeros of 
analytic functions. 
It involves the corner blocks of the inverse of the associated 
Hamiltonian matrix. As an illustration, numerical evaluations 
of the counting function of quasi 1D Anderson model are shown. 
\end{abstract}
\maketitle
\section{Introduction}
Consider the equation 
\begin{equation}
C_n u_{n-1}+A_nu_n +B_nu_{n+1}= E u_n \, , \qquad  n\in \mathbb{Z}
\label{equation}\end{equation}
where $A_n$, $B_n$, and $C_n$ are given complex non singular
square matrices of size $m$, $E$ is a parameter, and the vectors 
$u_n\in \mathbb{C}^m$ are unknown.\\
The case $A_n=A_n^\dagger$, $C_n=B_{n-1}^\dagger$ will be referred to as the
{\em Hermitian case}, and is common occurrence in physics: it describes a 
chain of $m-$level {\em atoms} with
Hamiltonians $A_n$, and couplings $B_n$ between neighbor atoms. In a 
lattice model for transport or in a model for a crystal,
the atoms may be thought of as sections of the lattice or crystal, 
each containing $m$ sites. In linear algebra, if off diagonal blocks 
are triangular, the equation represents the eigenvalue equation for a band 
matrix of bandwidth $2m+1$.\\
At each $n$ the equation provides $u_{n+1}$ in terms 
of $u_n$ and $u_{n-1}$, and an iterative evaluation of them can be started 
from initial conditions $u_0$, $u_1$. The process can be formulated via a 
$2m\times 2m$ transfer matrix:
\begin{eqnarray}
\left [ \begin{array}{c} u_{n+1}\\ u_n\end{array}\right ] = T(E)
\left [ \begin{array}{c} u_1 \\ u_0\end{array}\right ]
\label{transfer}
\end{eqnarray}
The $n$-step transfer matrix $T(E)$ is the product 
$t_n(E)t_{n-1}(E)\cdots t_1(E)$ of 1-step transfer matrices
with the block structure ($I_m$ is the unit matrix of size $m$)
\begin{equation}
t_k(E)=\left [\begin{array}{cc} B_k^{-1}(E-A_k) & -B_k^{-1}C_k^\dagger\\ 
I_m & 0 \end{array}\right ].
\end{equation}
We are concerned with the $2m$ exponents of the transfer matrix $T(E)$,
\begin{equation}
\xi_a = \frac{1}{n}\log |z_a|,
\end{equation} 
and their (normalized) counting function:
\begin{eqnarray}
\mathcal N (\xi)\,= \,\frac{1}{2m}
\{\, \# \,\xi_a:\,\xi_a<\xi \,\} =\frac{1}{2m} \sum_{a=1}^{2m} 
\theta (\xi-\xi_a)
\end{eqnarray}
An operative approach to the numerical evaluation of the exponents
is found in ref.\cite{Slevin}.
In several cases of interest, the exponents have definite large $n$ limits
(Lyapunov exponents) 
and describe the exponential rates of growth or decay in $n$ of the 
eigenvalues $z_a$, and of the solutions $u_n$ of the recursive 
eq.(\ref{equation}).
They are important, especially the largest or the (positive) 
smallest, in the study of dynamical systems, stability, localization 
problems, wave transmission in layered structures \cite{Crisanti}.
Analytic expressions for the distribution of Lyapunov exponents are 
known only for few cases, and such cases do not originate from a
Hamiltonian, i.e. a recursion equation like (\ref{equation}) 
\cite{Isopi,Gamba,Beenakker}. Most of the results where an Hamiltonian
is given first, were obtained numerically \cite{Kottos,Markos,Zhang} 
or via some expansion \cite{Derrida87,Baldes}.\\
The main result of the paper is the following exact formula, valid for a 
{\em single} transfer matrix of type (\ref{transfer}):
\begin{prop}[Counting function]
\begin{eqnarray}
&&{\mathcal N}(\xi) = \frac{1}{2}+ \left(\frac{2 \pi}{n}\right)^{-1}
\int_0^{\frac{2\pi}{n}} d\varphi\,
\frac{1}{2m}{\rm tr} \left [ z\,G^B (E,z)_{1n}B_n - 
\frac{1}{z} \,  G^B(E,z)_{n1}C_1\right ], \\ 
&&z=e^{\xi+i\varphi}\nonumber
\label{counting}
\end{eqnarray}
$G^B(E,z)_{ab}$ are the blocks of size $m$ of the resolvent 
$G^B(E,z)=[H^B (z)-E]^{-1}$, and $H^B(z)$ is the block tridiagonal matrix 
of size $mn$ with corners
\begin{equation}
H^B(z)=\left[\begin{array}{cccc}
A_1 & zB_1 & {} &  C_1/z \\
C_2/z  & \ddots & \ddots & {}\\
{} & \ddots & \ddots & zB_{n-1}\\
z B_n & {} & C_n/z & A_n 
\end{array}\right]
\end{equation}
\end{prop}
The suffix $B$ stands for ``balanced'', as this matrix is more tractable in
numerical calculations than the Hamiltonian matrix $H(z)$, to be 
introduced next.\\
Notes: 1) the integral will be shown to be a {\em contour} integral, 
and cancels all $z$ dependent terms of the Laurent expansion. 
Therefore one cannot remove the integral in the large $n$ limit. 2) The 
imaginary part of the integral is zero.\\ 
An expression of the counting function where the dependence on $z$ is made 
``explicit'' will be given in the Appendix, prop.~(\ref{explicit}).\\

\begin{figure}
\begin{center}
\includegraphics[angle=270,width=10cm]{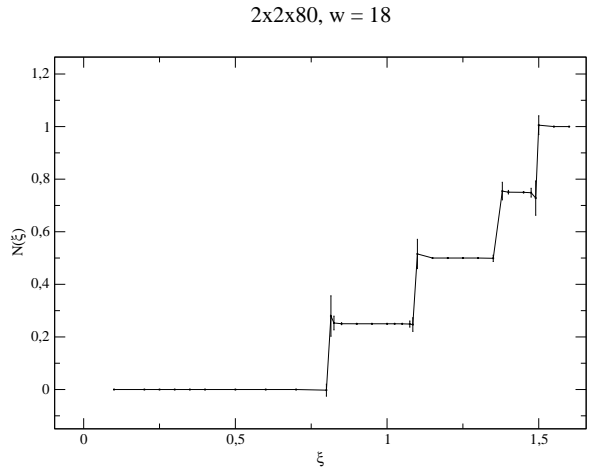}
\caption{\label{}The counting function $\mathcal N(\xi)$ (a different
normalization is used) for the Anderson model in a cube $2\times 2\times 80$, 
with disorder strength $w=18$, energy $E=0$. 
Each discontinuity marks a positive exponent $\xi_1,\ldots,\xi_4$ (the 
negative exponents are just their opposites).}
\end{center}
\end{figure}

\section{A spectral duality}
According to the {\em Argument principle} of complex analyis, the zeros 
of an analytic function $f(z)$ contained in the disk $|z|<R$ 
(inside the domain of analitycity) are enumerated by the following complex 
integral on a circle of radius $R$:
\begin{equation}
 n (R)  = \oint_{|z|=R}\frac{dz}{2\pi i} 
\frac{f'(z)}{f(z)} \label{euler}
\end{equation}
We are interested in the zeros of $f(z)=\det [T(E)-z]$, and their 
distribution ${\mathcal N}(\xi ) = (1/2m) n(e^{n\xi})$.
The result for the counting function rests on a remarkable 
duality that relates $f(z)$ to the characteristic polynomial of a 
Hamiltonian matrix $H(z)$, which naturally arises from the 
problem (\ref{equation}) restricted to $n$ steps in a ring topology, but
with generalized boundary conditions specified by a complex parameter $z$:
\begin{equation}
H (z)=\left[\begin{array}{cccc}
A_1 & B_1 & {} &  C_1/z \\
C_2  & \ddots & \ddots & {}\\
{} & \ddots & \ddots & B_{n-1}\\
z B_n & {} & C_n & A_n 
\end{array}\right]
\end{equation}

The polynomial $f(z)=\det [T(E) -z]$ has degree $2m$ in the variable $z$, 
and degree $nm$ in the parameter $E$. Duality states that the  
polynomial is proportional to 
the characteristic polynomial of the matrix $H(z)$:
\begin{prop}[Duality, \cite{Molinari03,Molinari08}]
\begin{equation}
\det [T(E)-z]= (-z)^m  \frac{\det [E-H(z)]}{\det[B_1\cdots B_n]}
\end{equation}
\end{prop}
It is a \emph{duality relation} as it exchanges the roles of 
the parameters $z$ and $E$ among matrices: {\em whenever $z$ is an
eigenvalue of the transfer matrix evaluated at $E$, the value $E$ is an 
eigenvalue of the block tridiagonal matrix $H(z)$}.
\begin{prop}
The matrix $H^B(z)$ is similar to $H(z^n)$: $H (z^n)\,=\,S(z) H^B(z) S(z)^{-1}$,
where $S(z)$ is the block diagonal matrix with blocks 
$S(z)_{ab}=\delta_{ab}z^a I_m$. 
\end{prop}  
\noindent
As a consequence the two matrices have the same eigenvalues. Another 
consequence is the property
$H^B  (ze^{-i2\pi/n})=S(e^{i2\pi/n}) H^B(z) S(e^{-i2\pi/n})$.\\

Before deriving the expression for the counting function, we quote interesting 
and related analytic consequences of duality. Duality and Jensen's formula of
complex analysis provide an equation for the exponents $\xi_a$ of 
the transfer matrix:
\begin{prop}[\cite{Molinari03}]
\begin{eqnarray}
&&\frac{1}{2m}\sum_{a=1}^{2m} 
\left( |\xi_a-\xi|+\xi_a+\xi \right )\,-\,\xi \nonumber\\
&&\quad =\frac{1}{mn}\int_0^{2\pi}\frac{d\varphi}{2\pi}\log\left |
\det [H(e^{n\xi+i\varphi})-E] \right |
-\frac{1}{mn}\sum_{k=1}^n\log \left | \det\,B_k \right |\label{dual2}
\end{eqnarray}
\end{prop}

\begin{cor} For $\xi=0$, the formula gives the sum of the positive exponents:
\begin{eqnarray}
&&\frac{1}{m}\sum_{a=1}^{2m} \xi_a\, \theta(\xi_a)\nonumber\\
&&\quad=\frac{1}{mn}\int_0^{2\pi}\frac{d\varphi}{2\pi}\log\left |
\det [H(e^{i\varphi})-E]\right |
-\frac{1}{mn}\sum_{k=1}^n\log \left |\det\,B_k \right |
\label{thoulesslike}
\end{eqnarray}
\end{cor}

\begin{remark}
The sum of all the exponents is $\frac{1}{n}\log |\det T(E)|$. Then:
\begin{equation}
\sum_{a=1}^{2m}\xi_a = \frac{1}{n}\sum_{k=1}^n \left(
\log|\det C_k|-\log |\det B_k|\right )
\end{equation}
\end{remark}

\begin{remark}
If the transfer matrix is symplectic, i.e. there is a matrix $\Sigma $
such that $T(E)^\dagger \Sigma T(E)=\Sigma $, then the exponents come
in opposite pairs. 
\begin{proof} If $z$ is an eigenvalue of $T(E)$ with eigenvector $\theta$, 
then $\Sigma \theta$ is an eigenvector of $T(E)^\dagger$ with 
eigenvalue $1/z$ . This means that $1/\overline{z}$ 
is eigenvalue of $T(E)$. The
corresponding exponents are either both zero or opposite real numbers.
\end{proof}   
\end{remark}

Eq.(\ref{thoulesslike}) is exact and applies to any single transfer
matrix. However it is reminiscent of the formula for the sum of the
Lyapunov exponents of an ensemble of random transfer matrices, 
obtained by Herbert, Jones and Thouless for $2\times 2$ matrices, 
and by Kunz, Souillard, Lacroix \cite{Lacroix} for larger ones: 
$$\frac{1}{m}\sum_a\lambda_a (E)= 
\int dE' \rho (E') \log |E-E'|+ \text{const.} $$
This formula was obtained for the Hermitian case, but holds also in
non Hermitian tridiagonal problems \cite{Derrida,Goldsheid05}.
In eq.(\ref{thoulesslike}) the angular average on $\varphi$ 
replaces the ensemble average, that produces the ensemble density of 
eigenvalues. It seems that an ergodic property is at play for such systems.

\section{The counting function}
The expression (\ref{counting}) for the counting function can be obtained
straightforwardly from eq.(\ref{dual2}). Here we give a direct  
proof based on the famous formula eq.(\ref{euler}), the
duality relation, and the simple formula 
$$\frac{d}{dz} \det M(z) = \det M(z)\; {\rm tr}\, \left [
M(z)^{-1} \frac{d}{dz}M(z)\right ].$$

\begin{proof}
Let $G(E,z)=[H(z)-E]^{-1}$, $f(z)=\det [T(E)-z]$, and use duality: 
\begin{eqnarray} 
\frac{f'(z)}{f(z)} &=& \frac{\frac{d}{dz}\left[ z^m\det[ H(z)-E]\right ]}
{z^m\det [H(z)-E]}
= \frac{m}{z}+ {\rm tr}\left[ G(E,z)\frac{d}{dz} H(z)\right ]
\nonumber \\
&=& \frac{m}{z} +{\rm tr} \left[ G(E,z)_{1n}B_n-\frac{1}{z^2}G(E,z)_{n1}
C_1\right ]
\nonumber
\end{eqnarray}
The complex integral of the first term is $m$. Inclusion of the normalization 
factor gives
\begin{eqnarray*}
{\mathcal N}(\xi)&=&\frac{1}{2}+\frac{1}{2m}\oint_{|z|=e^{n\xi}}\frac{dz}{2\pi iz}
{\rm tr} \left[z\, G(E,z)_{1n}B_n-\frac{1}{z}G(E,z)_{n1}C_1\right ]\\
&=&\frac{1}{2}+\frac{1}{2m}\int_0^{2\pi}\frac{d\theta}{2\pi}
{\rm tr} \left[ z\,G(E,z)_{1n}B_n-\frac{1}{z}G(E,z)_{n1}C_1\right ]
\end{eqnarray*}
where $z=e^{n\xi+i\theta}$. Next we set $\theta = n\varphi $, then 
$z=e^{n(\xi+i\varphi )}$. We introduce the resolvent of $H^B (z)$ and note that 
$ G(E,w^n)_{ab}=w^{a-b}G^B(E,w)_{ab}$ by virtue of the similarity; in particular
it is: 
$$w^nG(E,w^n)_{1n}=wG^B(E,w)_{1n}, \qquad 
\frac{1}{w^n}G(E,w^n)_{n1} = \frac{1}{w}G^B(E,w)_{n1} $$ 
The final formula is obtained (after renaming $w$ as $z$). 
Since the counting function is real, only the real part of the integral is 
nonzero, and:
\begin{eqnarray}
0 = \int_0^{\frac{2\pi}{n}} \frac{d\varphi}{2\pi}\,\,
{\rm Im}\,\,{\rm tr} \left [ z\,G^B(E,z)_{1n}B_n - 
\frac{1}{z} \, G^B(E,z)_{n1}C_1\right ]  .
\end{eqnarray}
\end{proof}
\begin{remark} While the Argument Principle requires the circle radius to
be $e^{n\xi}$, the use of the balanced martix scales the radius to $e^\xi$
and makes the formula for the counting function useful for numerical studies.
\end{remark} 
In numerical calculations the angular integral is replaced by a sampling on 
different angles. As the radius $e^\xi$ crosses an eigenvalue 
$e^{\xi_a+i\varphi_a}$ the counting function jumps to a new plateau after a 
strong oscillations due to close sampling points:
\begin{equation*}
\frac{1}{e^{\xi+i\varphi}-e^{\xi_a+i\varphi_a}}\approx 
\frac{e^{\xi_a+i\varphi_a}}{(\xi-\xi_a)^2+(\varphi-\varphi_a)^2}
[(\xi-\xi_a)-i(\varphi-\varphi_a)]
\end{equation*}

The formula requires the inversion of the matrix $H^B$, and the explicit 
dependence on the interesting parameter $z$ is lost. In the Appendix we
show that this dependence can be made more explicit by means of the resolvent 
equation for the matrix with corners removed.

\begin{figure}
\begin{center}
\includegraphics[angle=270,width=10cm]{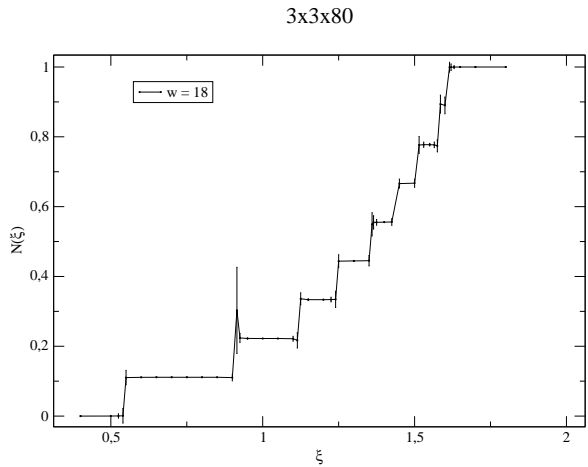}
\caption{\label{}The counting function $\mathcal N(\xi)$ (a different
normalization is used) for the Anderson model in a cube $3\times 3\times 80$, 
with disorder strength $w=18$, energy $E=0$. 
Each discontinuity marks a positive exponent $\xi_1,\ldots,\xi_9$.}
\end{center}
\end{figure}

\section{Anderson Model}
As an illustration of the formula, consider the Anderson model for 
the motion of a particle in a cubic lattice with the shape of a bar,
with a random potential. The particle may hop with equal amplitudes 
from a site $\vec k$ to neighboring ones $\vec k+\vec e$, and is subject 
to a site-potential $\epsilon (\vec k)$ (a random number uniformly 
distributed in the interval $[-w/2, w/2]$): 
\begin{equation*}
\sum_{\vec e} \psi (\vec k+\vec e) + \epsilon (\vec k)\psi (\vec k) = 
E \psi (\vec k)
\end{equation*}
If the bar is sliced perpendicularly to the long direction $(k=k_z)$, 
each slice is a square 2D lattice, with $m$ sites. Accordingly the 
Hamiltonian gains a block structure, with diagonal blocks $A_k$ describing 
slices, and blocks $B_k=C_k=I_m$ describing hopping among neighboring slices. 
The Schr\"odinger equation is now written as in (\ref{equation}):
$u_{k+1}+ u_{k-1} +A_k \,u_k = E\, u_k $,
where the $m$ components of $u_k$ are the amplitudes $\psi (\vec k)$ in 
the slice $k$.\\
Because of disorder, in the large $n$ limit, eigenstates are exponentially 
localized (Anderson localization) on a distance given by the inverse of 
the smallest positive exponent. In the large $n$ limit all the exponents 
$\pm\xi_a(E)$ exist, and are independent of the realization of the disorder.\\
The two figures show the counting function for a single realization of 
disorder, and a sample of length $80$ (cross section $2\times 2$, or 
$3\times 3$). Note the decrease of distances among exponents.\\
The subject on Anderson localization is vast. 
A discussion of the spectral properties of Anderson matrices with non 
Hermitian boundary conditions (the $z$ and $1/z$ dependence)
is presented in \cite{Molinari09}. The one dimensional case was studied
by Goldsheid and Khoruzhenko (Hermitian case) \cite{Goldsheid00} and several
others, and by us in the non Hermitian case \cite{Molinari09b}.

\section{Appendix}
 Let $h$ be the matrix $H(z^n)$ with 
corners removed (then it is $z$-independent):
\begin{equation}
h=\left[\begin{array}{cccc}
A_1 & B_1 & {} &  {} \\
C_2  & \ddots & \ddots & {}\\
{} & \ddots & \ddots & B_{n-1}\\
{} & {} & C_n & A_n 
\end{array}\right]
\end{equation}
and let $g(E)=[h-E]^{-1}$ be the resovent matrix, with blocks $g_{ab}$.
Then:
\begin{prop}\label{explicit} 

\begin{eqnarray}
&&{\rm tr}\,\left[z\; G^B(E,z)_{1n}B_n -\frac{1}{z}\; G^B (E,z)_{n1}C_1\right ]\\
&&\qquad ={\rm tr}\left[ \begin{array}{cc} 
-z^nB_n g_{1n}-I_m & -z^nB_n g_{11}\\ 
\frac{1}{z^n}C_1g_{nn} & \frac{1}{z^n}C_1 g_{n1}+I_m 
\end{array}\right ]^{-1}\nonumber
\end{eqnarray}
\begin{proof}
$\Delta (z^n)=H (z^n)-h$ is a matrix whose 
non zero blocks are  $\Delta (z^n)_{1n}=z^{-n}C_1 $ and 
$\Delta(z^n)_{n1}=z^n B_n$. The matrix equation for the resolvent is
\begin{equation}
G (E,z^n) = g (E) - g(E)\Delta (z^n)G(E,z^n)\label{resolvent}
\end{equation}
The four equations for the blocks $a,b=1,n$ can be put in matrix form 
\begin{eqnarray}
\left[ \begin{array}{cc} G_{1n} & G_{11}\\ 
G_{nn} & G_{n1}\end{array}\right ]= 
\left[ \begin{array}{cc} g_{1n} & g_{11}\\ 
g_{nn} & g_{n1}\end{array}\right ] 
- \left[ \begin{array}{cc} g_{1n} & g_{11}\\ 
g_{nn} & g_{n1}\end{array}\right ]
\left[ \begin{array}{cc} z^nB_n & 0  \\ 
 0 & C_1/z^n\end{array}\right ]
\left[ \begin{array}{cc} G_{1n} & G_{11}\\ 
G_{nn} & G_{n1}\end{array}\right ]\nonumber
\end{eqnarray}
The blocks of $G(E,z^n)$ are expressed in terms of the blocks of $G^B(E,z)$, 
\begin{eqnarray}
\left[ \begin{array}{cc} G_{1n} & G_{11}\\ G_{nn} & G_{n1}
\end{array}\right ]= 
\left[ \begin{array}{cc} 1/z^n & 0\\ 0 & z^n \end{array}\right ] 
\left[ \begin{array}{cc} zG^B_{1n} & z^n G^B_{11}\\ 
G^B_{nn}/z^n & G^B_{n1}/z \end{array}\right ] \nonumber
\end{eqnarray}
Then
\begin{eqnarray}
&&\left[ \begin{array}{cc} zG^B_{1n} & z^n G^B_{11}\\ 
G^B_{nn}/z^n & G^B_{n1}/z \end{array}\right ] =
\left[ \begin{array}{cc} z^n g_{1n} & z^n g_{11}\\ 
g_{nn}/z^n & g_{n1}/z^n \end{array}\right ] \nonumber \\
&&\quad -
\left[ \begin{array}{cc} z^n g_{1n} & z^n g_{11}\\ 
g_{nn}/z^n & g_{n1}/z^n \end{array}\right ] 
\left[ \begin{array}{cc} B_n & 0\\ 
0 & C_1 \end{array}\right ]
\left[ \begin{array}{cc} zG^B_{1n} & z^n G^B_{11}\\ 
G^B_{nn}/z^n & G^B_{n1}/z \end{array}\right ]\nonumber
\end{eqnarray}
with solution
\begin{eqnarray}
\left[ \begin{array}{cc} zG^B_{1n} & z^n G^B_{11}\\
G^B_{nn}/z^n & G^B_{n1}/z \end{array}\right ] 
=\left( 
\left[ \begin{array}{cc} g_{1n} & g_{11}\\ 
g_{nn} & g_{n1} \end{array}\right ]^{-1}
\left[ \begin{array}{cc} 1/z^n & 0\\ 
0 & z^n \end{array} \right ] +
\left[ \begin{array}{cc} B_n & 0\\ 
0 & C_1\end{array} \right ]\right)^{-1} \nonumber
\end{eqnarray}
Finally, left multiplication by the diagonal matrix 
with blocks $B_n$ and $-C_1$
gives
\begin{eqnarray}
&&\left[ \begin{array}{cc} zB_nG^B_{1n} & z^nB_n G^B_{11}\\ 
-C_1G^B_{nn}/z^n & -C_1G^B_{n1}/z \end{array}\right ] \nonumber\\ 
&&\quad =\left( 
\left[ \begin{array}{cc} g_{1n} & g_{11}\\ 
g_{nn} & g_{n1} \end{array}\right ]^{-1}
\left[ \begin{array}{cc} B_n^{-1}/z^n& 0\\ 
0 & -z^nC_1^{-1}\end{array} \right ] +
\left[ \begin{array}{cc} I_m & 0\\ 
0 & -I_m\end{array} \right ]\right)^{-1} \nonumber
\end{eqnarray}
The r.h.s. is simplified by means of the following identity
for invertible matrices $A$ and $B$: $(A^{-1}+B^{-1})^{-1}=A-A(A+B)^{-1}A$,
\begin{eqnarray}
&&=\left[ \begin{array}{cc} I_m & 0\\ 0 & -I_m\end{array} \right ]
-\left[ \begin{array}{cc} I_m & 0\\ 0 & -I_m\end{array} \right ]\nonumber\\
&& \qquad \times \left(
\left[ \begin{array}{cc} z^nB_n & 0\\ 
0 & -\frac{1}{z^n}C_1\end{array} \right ]
\left[ \begin{array}{cc} g_{1n} & g_{11}\\ 
g_{nn} & g_{n1} \end{array}\right ]
 +
\left[ \begin{array}{cc} I_m & 0\\ 
0 & -I_m\end{array} \right ]\right)^{-1} 
\left[ \begin{array}{cc} I_m & 0\\ 
0 & -I_m\end{array} \right ]\nonumber
\end{eqnarray}
The trace is then taken:
\begin{eqnarray}
\quad{\rm tr}\left[ zB_nG^B_{1n} -\frac{1}{z} C_1G^B_{n1} \right ]=
{\rm tr}\left[ \begin{array}{cc} 
-z^nB_n g_{1n}-I_m & -z^nB_n g_{11}\\ 
\frac{1}{z^n}C_1g_{nn} & \frac{1}{z^n}C_1 g_{n1}+I_m 
\end{array}\right ]^{-1}\label{matrixprop}
\end{eqnarray}
As a further step one could perform the the block inversion of the matrix 
by Schur's formula. Since the trace is then taken, only the diagonal 
blocks are needed: 
\begin{eqnarray}
&& =-{\rm tr}\left[ z^n B_n g_{1n}+ I_m+
B_n g_{11}(\frac{1}{z^n}C_1 g_{n1}+ I_m)^{-1}C_1g_{nn} \right]^{-1} \nonumber \\
&&\qquad
+ \,{\rm tr}\left[\frac{1}{z^n}C_1g_{n1}+ I_m + C_1 g_{nn}(z^nB_n g_{1n}+I_m
)^{-1}B_ng_{11} \right]^{-1}
\end{eqnarray}
\end{proof}
\end{prop}
This final expression allows in principle the evaluation of the counting
function with a single big matrix inversion for obtaining the four 
corner matrix blocks $g_{ab}$ ($a,b=1,n$). Then, for each $\xi$ value, 
the inversion of a square $m\times m$ matrix is required.

%\begin{figure}%
%\begin{center}
%\includegraphics[width=10cm]{plot_id_n4_4_40_Wvari.eps}%\hspace{0.3cm}
%%\includegraphics[width=2cm]{nhtrid_n800_xi1_nangle1.ps}
%%\includegraphics[width=6.5cm]{plot_xi.eps}
%\caption{\label{}The normalized counting function $\mathcal N(\xi)$ 
%for the Anderson model in a cube $4\times 4\times 40$, $w=8,18,24$, $E=0$. 
%Each discontinuity marks an exponent $\xi_a$, $a=1...16$.}
%\end{center}
%\end{figure}

\vfill
\end{document}